\documentclass[aps,prb,twocolumn,superscriptaddress,showpacs,preprintnumbers]{revtex4-1}
\usepackage{graphicx }
\usepackage{dcolumn}
\usepackage{bm}
\usepackage{braket}
\usepackage{amsmath}
\usepackage[caption=false]{subfig} 
\usepackage{setspace} 
\usepackage{tabularx}
\usepackage{color}
\usepackage{epstopdf}
\newcolumntype{Y}{>{\centering\arraybackslash}X}

\newcommand{\vzz}{\textit{V}$_{zz}$}
\newcommand{\td}{$\theta_{\mathrm{M}}$}
\newcommand{\tn}{\textit{T}$_{\mathrm{N}}$}

\newcommand{\ton}{\textit{T}$_{\mathrm{N}}^{\mathrm{onset}}$}
\newcommand{\tfull}{\textit{T}$_{\mathrm{N}}^{\mathrm{100 \%}}$}
\newcommand{\bhf}{\textit{B}$_{\mathrm{hf}}$}
\newcommand{\bc}{$\beta_{\mathrm{c}}$}
\begin{document}


\title{Suppression of the magnetic order in CeFeAsO: non-equivalence of hydrostatic and chemical pressure}
\author{Philipp Materne}
\altaffiliation[]{pmaterne@anl.gov}
\affiliation{Argonne National Laboratory, Lemont, IL 60439, USA}
\affiliation{Institute of Solid State and Materials Physics, TU Dresden, D-01069 Dresden, Germany}
\author{Wenli Bi}
\affiliation{Argonne National Laboratory, Lemont, IL 60439, USA}
\affiliation{Department of Geology, University of Illinois at Urbana-Champaign, Urbana, Illinois 61801, USA}
\author{Esen Ercan Alp}
\author{Jiyong Zhao}
\author{Michael Yu Hu}
\affiliation{Argonne National Laboratory, Lemont, IL 60439, USA}
\author{Anton Jesche}
\affiliation{EP VI, Center for Electronic Correlations and Magnetism, Institute of Physics, University of Augsburg, D-86159 Augsburg, Germany}
\author{Christoph Geibel}
\affiliation{Max Planck Institute for Chemical Physics of Solids, N\"othnitzer Str. 40, 01187 Dresden, Germany}
\author{Rhea Kappenberger}
\affiliation{Leibniz Institute for Solid State and Materials Research (IFW) Dresden, D-01069, Germany}
\affiliation{Institute of Solid State and Materials Physics, TU Dresden, D-01069 Dresden, Germany}
\author{Saicharan Aswartham}
\affiliation{Leibniz Institute for Solid State and Materials Research (IFW) Dresden, D-01069, Germany}
\author{Sabine Wurmehl}
\author{Bernd B\"uchner}
\affiliation{Leibniz Institute for Solid State and Materials Research (IFW) Dresden, D-01069, Germany}
\affiliation{Institute of Solid State and Materials Physics, TU Dresden, D-01069 Dresden, Germany}
\author{Dongzhou Zhang}
\affiliation{Hawaii Institute of Geophysics and Planetology, School of Ocean and Earth Science and Technology, University of Hawaii at Manoa, Honolulu, HI 96822, USA}
\author{Til Goltz}
\author{Johannes Spehling}
\author{Hans-Henning Klauss}
\affiliation{Institute of Solid State and Materials Physics, TU Dresden, D-01069 Dresden, Germany}


\date{\today}

\begin{abstract}
We present a detailed investigation of the electronic properties of $\mathrm{CeFeAsO}$ under chemical (As by P substitution) and hydrostatic pressure by means of in-house and synchrotron M\"ossbauer spectroscopy.
The Fe magnetism is suppressed due to both pressures and no magnetic order was observed above a P-substitution level of 40\;{}\% or 5.2\;{}GPa hydrostatic pressure.
We compared both pressures and found that the isovalent As by P substitution change the crystallographic and electronic properties differently than hydrostatic pressure.

\end{abstract}

\maketitle
\section{introduction}
The parent compounds of the 122 and 1111 families of the iron-based superconductors show spin density wave (SDW) order below the magnetic transition temperature \tn.\cite{PhysRevB.78.020503,PhysRevLett.101.077005}
By changing a non-temperature control parameter the SDW order can be suppressed.
These control parameters can be classified in the following way: i) electron doping (Fe$\rightarrow$Co,\cite{PhysRevB.83.094523,0295-5075-91-4-47008} O$\rightarrow$F\cite{nmat2397}), ii) hole doping (Ca$\rightarrow$Na,\cite{PhysRevB.92.134511} Ba$\rightarrow$K\cite{PhysRevLett.107.237001}), iii) isovalent substitution (As$\rightarrow$P\cite{1742-6596-551-1-012025, 0953-8984-21-38-382203,PhysRevB.86.094521}), and iv) external pressure.\cite{PhysRevB.79.224518, 0953-8984-21-1-012208,JPSJ.77.113712}
Both electron and hole doping change the amount of conduction electrons.
The nominal valence electron count remains constant in the case of isovalent substitution and external pressure.
The isovalent substitution of As by P results in chemical pressure due to the substitution of a larger by a smaller atom.
The resulting question is: what are the differences between chemical and hydrostatic pressure?
It was shown that both methods of achieving pressures result in a similar suppression of the magnetic order.\cite{PhysRevB.82.014513, nmat2443, PhysRevB.79.224518,e2010-10522-1, PhysRevB.79.172506, JPSJ.79.123706, 0295-5075-87-1-17004, 0295-5075-86-4-47002, JPSJ.77.113712, JPSJS.77SC.78}

The CeFeAsO system is of particular interest due to the interaction of the Fe 3\textit{d} and Ce 4\textit{f} electrons.
CeFeAsO shows spin density wave order of the Fe 3\textit{d} electrons below $\sim$ 145 K and antiferromagnetic order of the Ce 4\textit{f} electrons below $\sim$ 3.7 GPa.\cite{PhysRevB.86.020501}
A strong Ce-Fe coupling at temperatures much higher than the Ce magnetic ordering temperature was found.\cite{PhysRevB.80.094524}
Upon P substitution the Fe magnetic order is suppressed and no Fe magnetic order was observed for \textit{x} $\geq$ 37 \%.\cite{PhysRevB.86.020501, PhysRevLett.104.017204}
In contrast, the Ce magnetic ordering temperature remains constant for \textit{x} $<$ 30 \%.\cite{PhysRevB.86.020501}
For \textit{x} $\geq$ 30 \% the Ce magnetic order changes from antiferromagnetic to ferromagnetic.\cite{PhysRevB.86.020501}
Superconductivity was observed for \textit{x} $\sim$ 30 \%.\cite{PhysRevB.86.020501}
Resistivity measurements have shown the absence of superconductivity in CeFeAsO up to 50 GPa.\cite{PhysRevB.83.094528}
The application of hydrostatic pressure on P substituted samples indicated that hydrostatic pressure and P substitution change the electronic structure differently.\cite{PhysRevB.86.134523}
However, to resolve the microscopic changes in the magnetic structure a local probe is needed.

We studied the electronic hyperfine parameters as a function of P substitution and hydrostatic pressure in $\mathrm{CeFeAsO}$ by means of in-house and synchrotron M\"ossbauer spectroscopy as well as x-ray diffraction.
We found a quantitatively different behavior for P substitution and hydrostatic pressure.

The work is organized in the following way: the experimental details will be presented in Sec. \ref{sec:experimental} and the obtained results in Sec. \ref{sec:xrd} and \ref{sec:results}.
The discussion of our results is given in Sec. \ref{sec:discussion} followed by a summary and conclusion in Sec. \ref{sec:summary}.

\section{experimental details}
	\label{sec:experimental}
	
Powder samples of CeFeAs$_{1-x}$P$_x$O with \textit{x}\;{}=\;{}0, 5, 15, 22, 30, 35, 40, 90, and 100\;{}\% were investigated by in-house M\"ossbauer spectroscopy at the Institute of Solid State and Material Physics, TU Dresden, Germany.
The P-substitution level \textit{x} is given in nominal values.\cite{PhysRevB.81.134525}
M\"ossbauer spectra were recorded at temperatures between 1.8 and 305\;{}K using a CryoVac Konti IT cryostat in standard transmission geometry.
As a $\gamma$ source $^{57}$Co in a rhodium matrix was used with an emission line width (HWHM) of 0.135(5)\;{}mm/s.
Isomer shifts are given with respect to $\alpha$-Fe at room temperature.
Powder samples were homogeneously distributed in thin polyamide PA6.6 sample holders of 13\;{}mm diameter.
The sample synthesis of the CeFeAs$_{1-x}$P$_x$O powder is described elsewhere.\cite{PhysRevB.81.134525}
The in-house M\"ossbauer spectra were analyzed using the M\"ossfit software.\cite{kamusella2016free}

Single crystals of $\mathrm{CeFeAsO}$ were investigated via time-domain synchrotron M\"ossbauer spectroscopy (SMS), also known as nuclear forward scattering, at the beamline 3ID-B of the Advanced Photon Source (APS) at Argonne National Laboratory, USA.
The experiments were performed in the hybrid operation mode which allows the high precision measurement of hyperfine interactions by offering a time window for data collection of 1.5\;{}$\mu$s.
The single crystals were enriched to an abundance of 10\;{}\% $^{57}$Fe.
They were grown similar to LaFeAsO\cite{Kappenberger20189} and characterized by energy-dispersive x-ray spectroscopy and x-ray diffraction (XRD).
SMS spectra were recorded at temperatures between 10 and 150\;{}K and at pressures between 0.5 and 14\;{}GPa using a special He-flow miniature cryostat and a diamond anvil cell.\cite{Bihf5283,1.4999787}
For pressures up to 6\;{}GPa diamond anvils with 800\;{}$\mu$m culet size and for higher pressures diamond anvils with 500\;{}$\mu$m culet size were used.
Pressures were changed at low temperatures through a gas membrane.
The pressure was measured \textit{in situ} by an online ruby system.
A Re gasket was pre-indented to a thickness of 80\;{}$\mu$m (140\;{}$\mu$m) and a 250\;{}$\mu$m (400\;{}$\mu$m) hole was electro-sparked to act  as the sample chamber for the 500\;{}$\mu$m (800\;{}$\mu$m) diamond anvils.
As the pressure transmitting medium Ne and a 4:1 mixture of Methanol and Ethanol were used to ensure hydrostaticity.
The uncertainty in the pressure determination is 0.1\;{}GPa if not stated otherwise.
Single crystals of 50$\times$50$\times$45\;{}$\mu$m$^3$ and 130$\times$130$\times$25\;{}$\mu$m$^3$ for the 500\;{}$\mu$m and 800\;{}$\mu$m diamond anvils were used, respectively.
The single crystals were aligned with the crystallographic \textit{ab}-axis perpendicular to the incident beam.
The beam size was 10$\times$15\;{}$\mu$m$^2$ (FWHM).
The SMS spectra were analyzed using the CONUSS software.\cite{CONUSS}
Both M\"ossfit and CONUSS exactly diagonalize the hyperfine Hamiltonian taking into account both electric and magnetic hyperfine interactions.
For the former the transmission integral formalism and for the latter the thin absorber approximation was used.
XRD experiments were conducted at the 13BM-C beamline of the APS using $\mathrm{CeFeAsO}$-powder at room temperature.\cite{jove-119-54660}
We used x-rays with a wavelength of 0.434\;{}\AA, a Re gasket as described earlier and Daphne oil 7575 as the pressure transmitting medium.
The x-ray diffraction patterns were analyzed using GSAS-II.\cite{Tobyaj5212}
The numerical results of our XRD study are recorded in the supplement together with the numerical values of chosen diagrams.


\section{X-ray diffraction results}
	\label{sec:xrd}

\begin{figure}[htbp]
	\centering
	\includegraphics[width= \columnwidth]{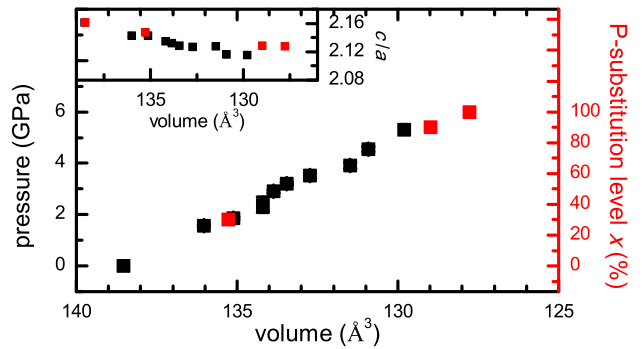} 	
	\caption{Unit cell volume at room temperature as a function of applied pressure \textit{p} (black) and the P-substitution level \textit{x} (red, taken from Ref.\cite{AJesche_Diss}). 100\;{}\% P-substitution ($\mathrm{CeFeAsO}$ $\rightarrow$ CeFePO) reduces the unit cell volume similar to the application of 6\;{}GPa hydrostatic pressure.}
	\label{fig:xrd}
\end{figure}

To compare the structural effects of P substitution and external pressure we performed XRD measurements up to 5.3\;{}GPa at room temperature using a $\mathrm{CeFeAsO}$ powder sample.
The resulting unit cell volumes and \textit{c}/\textit{a} ratios are shown in Fig.~\ref{fig:xrd} together with published data on CeFeAs$_{1-x}$P$_x$O for comparison.\cite{AJesche_Diss}
At room temperature and ambient pressure $\mathrm{CeFeAsO}$ crystallizes in a tetragonal structure with the space group \textit{P}4/\textit{nmm}.\cite{1367-2630-11-10-103050}
We found no indications for structural transitions up to 5.3\;{}GPa and that the \textit{c}/\textit{a} ratio is more reduced in the case of hydrostatic pressure than upon P substitution.
By comparing the unit cell volumes we found that 100\;{}\% P-substitution ($\mathrm{CeFeAsO}$ $\rightarrow$ CeFePO) has the same effect as the application of 6\;{}GPa hydrostatic pressure.

\begin{figure}[htbp]
	\centering
	\includegraphics[width= \columnwidth]{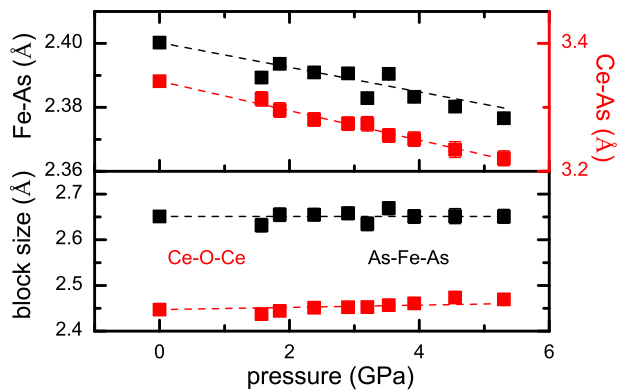} 	
	\caption{Atomic distances and block sizes for CeFeAsO as a function of pressure.}
	\label{fig:lattice}
\end{figure}

Atomic distances and block sizes are shown in Fig. \ref{fig:lattice}.
We found that the Ce-O-Ce as well as the As-Fe-As block size are pressure independent.
Both the Fe-As and Ce-As distances are reduced with increasing hydrostatic pressure.
Therefore the reduction in the unit cell volume is achieved by reducing the distance between the Ce-O-Ce and As-Fe-As blocks.
In contrast the unit cell compression due to the P-substitution is caused by a compression of the As-Fe-As layer.\cite{PhysRevLett.104.017204}

\section{M\"ossbauer spectroscopy results}
	\label{sec:results}
	
\begin{figure*}[htbp]
	\centering
	\includegraphics[]{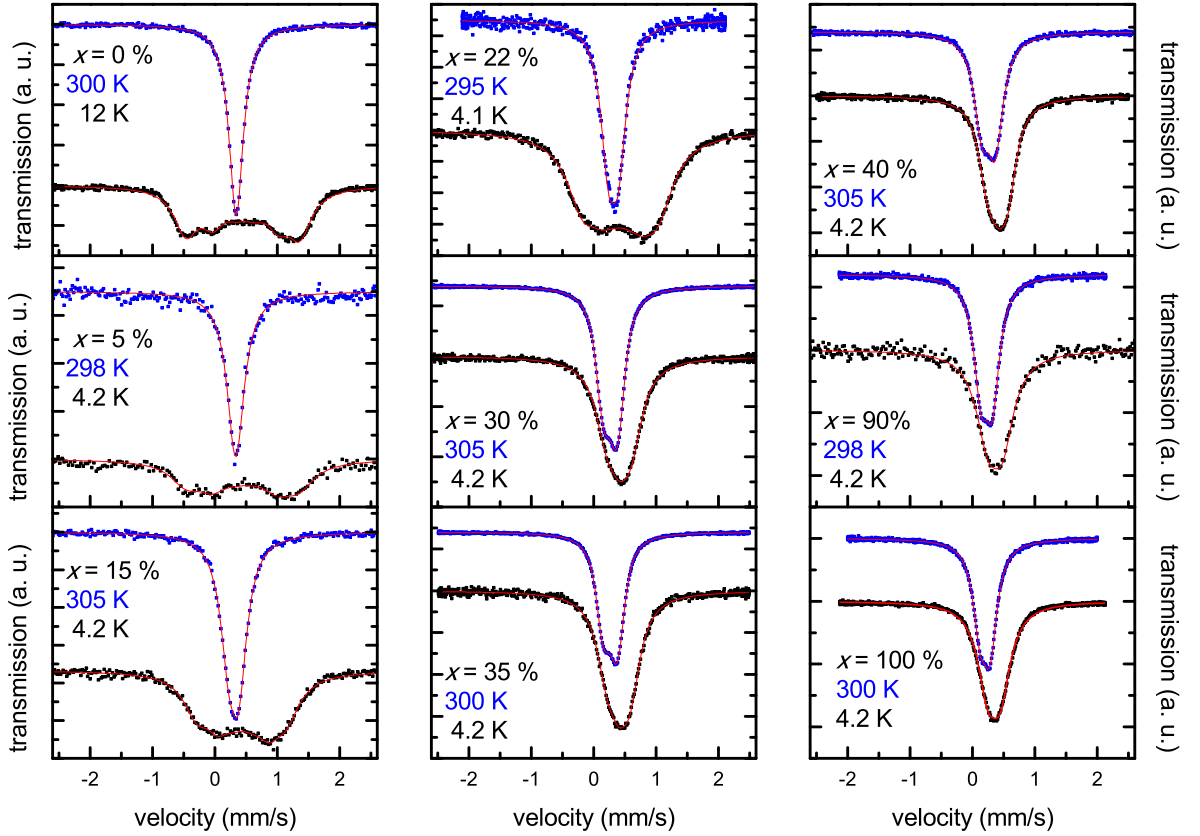} 
	\caption{M\"ossbauer spectra of CeFeAs$_{1-x}$P$_x$O in the paramagnetic and magnetically ordered phase. The solid red lines are theoretically calculated spectra. See text for details.}
	\label{fig:CeFeAsPO-spectra}
\end{figure*}

M\"ossbauer spectra of CeFeAs$_{1-x}$P$_x$O in the paramagnetic and magnetically ordered phase are shown in Fig.~\ref{fig:CeFeAsPO-spectra}.
In the paramagnetic phase for 0\;{}$\leq$\textit{x}\;{}$\leq$\;{}22\;{}\% a non-resolved doublet structure and for 30\;{}$\leq$\textit{x}\;{}$\leq$\;{}100\;{}\% an asymmetric doublet structure is observed.
For powder samples one would expect a symmetric spectrum as the angle between the incident $\gamma$ and the principal axis of the electric field gradient (EFG) is averaged out.
The asymmetric paramagnetic spectra for \textit{x}\;{}$\geq$\;{}30\;{}\% indicate that the samples consists of tiny polycrystalline platelets instead of powder in accordance with the plate-like crystal habit.
As a consequence the angle between the incident $\gamma$ and the principal axis of the EFG is not averaged out resulting in an asymmetric doublet.
The magnetically ordered phase is characterized by a sextet structure for \textit{x}\;{}$\leq$\;{}22\;{}\% while for \textit{x}\;{}=\;{}30 and 35\;{}\% a broadening and a symmetrization of the spectra was observed.

\begin{figure*}[htbp]
	\centering
	\includegraphics[]{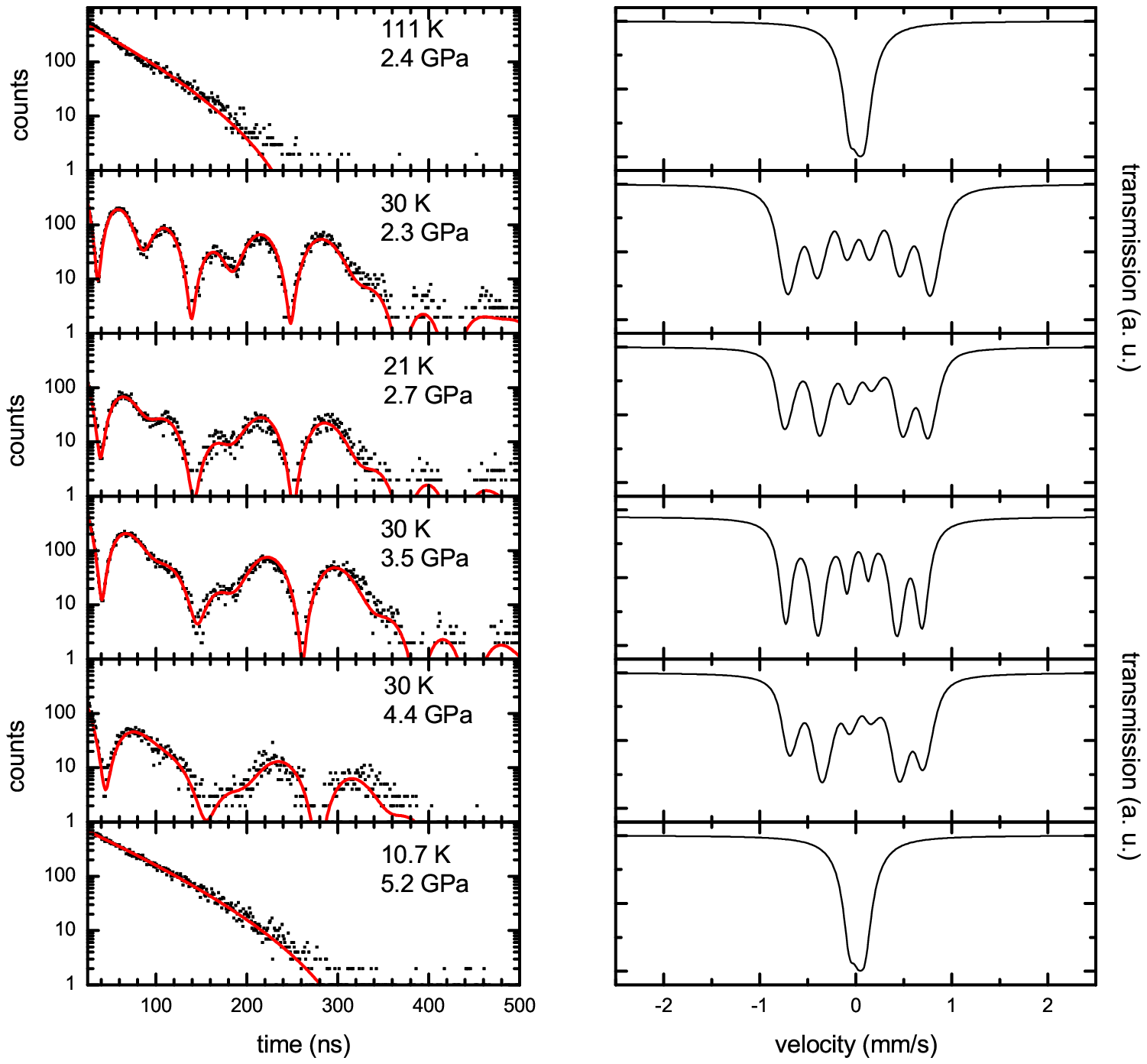} 	
	\caption{Synchrotron M\"ossbauer spectroscopy spectra of $\mathrm{CeFeAsO}$ in the paramagnetic and magnetically ordered phase for various pressures (left column). The solid red lines are theoretical spectra. The corresponding spectra of the fit in the energy domain are shown in the right column for clarity. See text for details.}
	\label{fig:spectra-pressure}
\end{figure*}

SMS spectra in the paramagnetic and magnetically ordered phase for various pressures are shown in Fig.~\ref{fig:spectra-pressure}.
In the paramagnetic phase no oscillations in the time spectra were observed up to pressures of 14~\;{}GPa.
The magnetically ordered phase is characterized by many oscillations with additional wiggles due to the angle between the magnetic hyperfine field and the incident $\gamma$ beam.

\subsection{Electric field gradient}
In the principal axis system, the EFG is fully determined by its \textit{z} component \vzz \,{} and the asymmetry parameter $\eta$.
The latter is zero due to the tetragonal symmetry of the crystallographic structure in the paramagnetic phase.
In the magnetically ordered phase no non-zero $\eta$ was observed which is consistent with the absence of an orthorhombic distortion.\cite{PhysRevB.86.020501}
Neutron scattering experiments report an orthorhombicity of 0.5\;{}\% for $\mathrm{CeFeAsO}$ which is suppressed due to P substitution.\cite{PhysRevLett.104.017204}
However, the changes in the EFG due to the orthorhombic distortion are below the resolution limit of our method.

In M\"ossbauer spectroscopy an energy shift, the so-called quadrupolar splitting \textit{QS}, due to the interaction of the Fe nucleus with an EFG rather than \vzz \,{} itself is measured.
From the \textit{QS} the electric field gradient \vzz\,{} at the Fe nucleus can be deduced.
Here we provide both \textit{QS} in mm/s and \vzz\,{} in V/\AA$^2$.
The conversion factor is 1\;{}V/\AA$^2$\;{}=\;{}$-$\;{}0.0167\;{}mm/s which corresponds to a nuclear quadrupole moment of Fe of 160\;{}mb.\cite{00268970802018367,PhysRevLett.75.3545}
At this point we want to emphasize that in the paramagnetic phase only the absolute value of \vzz\,{} is obtained.
However, it was shown that in the LaFeAsO-based compounds \vzz\,{} is positive.\cite{PSSB201600160,0805.0041}
In the magnetic phase we obtained a positive \vzz\,{} value and thus we are confident that this is also the case in the paramagnetic phase of the $\mathrm{CeFeAsO}$ series.

\begin{figure}[htbp]
	\centering
	\includegraphics[width= \columnwidth]{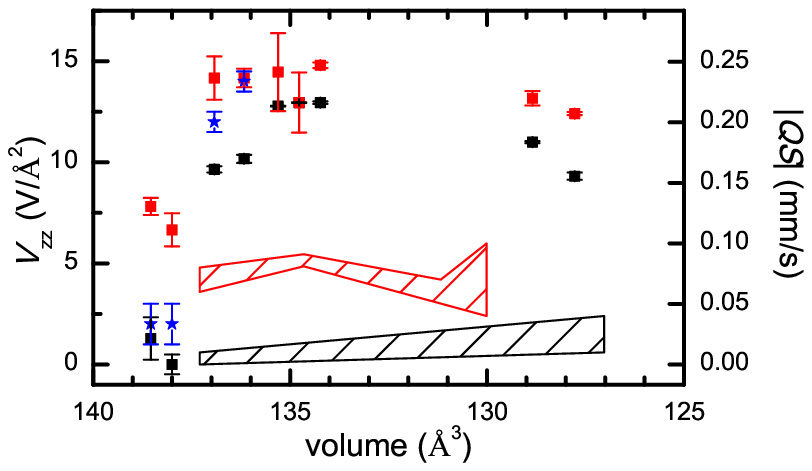} 	
	\caption{\vzz \,{} as a function of the P-substitution level \textit{x} (squares) and external pressure (shaded area) at the highest examined temperatures (295 to 305\;{}K, black), at the lowest examined temperatures (2 to 12\;{}K for P-substitution series, 12 to 30\;{}K for pressure measurements, red), and at \ton~ (blue) in CeFeAsO. At the highest measured temperatures in the paramagnetic phase \vzz \,{}is close to zero for $\mathrm{CeFeAsO}$ indicating a nearly spherical electron distribution around the Fe nucleus. \vzz \,{} shows a parabolic behavior as a function of \textit{x} with a maximum at intermediate \textit{x}. The increase from highest to lowest temperatures is largest for \textit{x}\;{}$\leq$\;{}5\;{}\% and decreases with increased \textit{x}.}
	\label{fig:Vzz}
\end{figure}

Experimentally determined \vzz \,{}values at various temperature regions between 2 and 305\;{}K are shown in Fig.~\ref{fig:Vzz}.
At room temperature \vzz \,{} is close to zero for $\mathrm{CeFeAsO}$ indicating a nearly spherical electron distribution around the Fe nucleus.
\vzz \,{} shows a parabolic behavior as a function of \textit{x} with a maximum at intermediate \textit{x}.
The \vzz \,{}values of $\mathrm{CeFeAsO}$ and CeFePO are equal to formerly reported data.\cite{1367-2630-11-2-025011, JPSJ.81.064714}

For \textit{x}\;{}$\leq$\;{}22\;{}\%, \vzz \,{} increases by $\approx\;{}2\;{}$V/\AA$^2$ between room temperature and the onset temperature of the magnetic order, \ton.
This increase of \vzz \,{} inside the paramagnetic phase as a function of temperature is likely a steric effect such as a change in the \textit{c}/\textit{a} ratio or the anion height.

At \ton, which we defined as the highest temperature with a non-zero magnetic volume fraction, \vzz \,{} jumps from 2(1)\;{}V/\AA$^2$ to 7(1)\;{}V/\AA$^2$ for \textit{x}\;{}=\;{}0 and 5\;{}\% and from 12.0(5)\;{}V/\AA$^2$ to 14.0(5)\;{}V/\AA$^2$ for \textit{x}\;{}=\;{}15 and 22\;{}\%, respectively.
This indicates a change of the electron distribution and hence of \vzz \,{} due to the magnetic phase transition.

In the magnetically ordered phase \vzz \,{} remains constant within error bars down to lowest measured temperatures.
The increase in \vzz \,{} at the magnetic phase transition is suppressed similar to the reduction of \ton~ and the magnetic hyperfine field as a function of \textit{x}.
No influence of the Ce magnetic order on \vzz \,{} has been observed similar to the unsubstituted compound.\cite{1367-2630-11-2-025011}
For \textit{x}\;{}$\geq$\;{}30\;{}\% \vzz \,{} increases upon cooling and saturates below 100\;{}K.

The SMS spectra in the paramagnetic phase show no oscillations up to 330\;{}ns. 
This gives an upper boundary for the absolute value of the quadrupole splitting of $\sim$\;{}0.1\;{}mm/s (6\;{}V/\AA$^2$). 
Analyzing the spectra gives a value of $<$\;{}0.01\;{}mm/s (0.6\;{}V/\AA$^2$) at 1\;{}GPa which is similar to the \vzz \,{}values within error bars at ambient conditions.
By increasing the external pressure, \vzz \,{} increases to 0.08(1)\;{}mm/s at 7\;{}GPa and 0.11(1)\;{}mm/s at 14\;{}GPa with both values obtained at 15\;{}K.

In the magnetically ordered phase, the quadrupole splitting jumps to 0.08(2)\;{}mm/s (4.8(1.2)\;{}V/\AA$^2$) and stays constant within error bars down to lowest temperatures.


\subsection{Magnetic order}

\begin{figure}[htbp]
	\centering
	\includegraphics[width= 8.6cm]{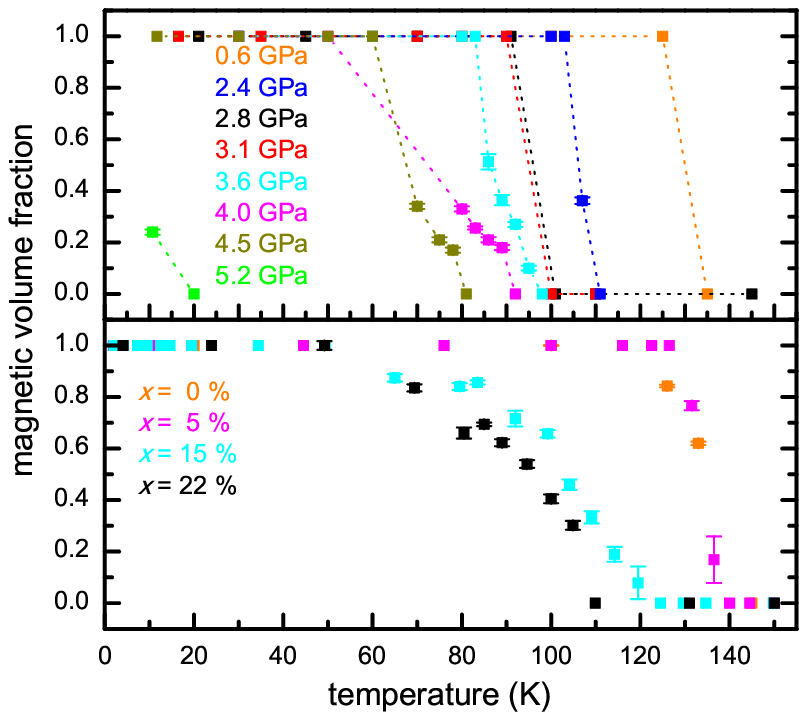} \label{fig:MVF-p-x}
	\caption{Magnetic volume fraction in $\mathrm{CeFeAsO}$ as a function of the applied pressure \textit{p} (top) and of the P-substitution level \textit{x} (bottom). The onset temperature of the magnetic phase transition is suppressed with increasing \textit{x} and \textit{p}. The phase transition region broadens for higher \textit{x} whereas it remains sharp for increasing pressure. Lines are guide to the eye.}
	\label{fig:magnetic volume fraction}
\end{figure}

The temperature dependence of the magnetic volume fraction for \textit{x}\;{}$\leq$\;{}22\;{}\% and applied pressures of \textit{p}\;{}$\leq$\;{}5.2\;{}GPa is shown in Fig.~\ref{fig:magnetic volume fraction}.

\ton~ decreases with increasing \textit{x} and \textit{p}.
The phase transition region, which can be defined as the temperature difference between \ton~ and \tfull (the highest temperature with a magnetic volume fraction of 100\;{}\%) increases from $\approx$\;{}10\;{}K for \textit{x}\;{}$\leq$\;{}5\;{}\% to 60(10)\;{}K for \textit{x}\;{}=\;{}15 and 22\;{}\%.
For \textit{x}\;{}=\;{}30 and 35\;{}\% no magnetic volume fraction was extracted as the obtained magnetic hyperfine fields are too small to distinguish between a) a smaller hyperfine field and 100\;{}\% magnetic volume fraction or b) a slightly larger hyperfine field and a magnetic volume fraction of $<$\;{}100\;{}\%, in particular as \vzz \,{} shows no measurable between the paramagnetic and magnetically ordered phase (in contrast to \textit{x}\;{}$\leq$\;{}22\;{}\%).
As a consequence, the magnetic volume fraction was set to 100\;{}\% in the magnetically ordered phase.
This may influence the analysis close to the phase transition temperature but the low-temperature behavior and therefore the saturated magnetic hyperfine field is unaffected.

\ton~ is reduced as a function of the the applied pressure consistent with reported results from electrical resistivity measurements.\cite{PhysRevB.83.094528}
The phase transition region between \ton~ and \tfull~ stays constant up to an applied pressure of at least 4.5\;{}GPa.
For an applied pressure of 5.2\;{}GPa we found a magnetic volume fraction of 24(1)\;{}\% while for 5.1\;{}GPa we observed a pure paramagnetic signal at lowest measured temperature.
Note that the given pressure values are determined at the ruby position with an uncertainty in the pressure determination of 0.1\;{}GPa.

\begin{figure}[htb]
	\centering
	\captionsetup[subfloat]{position=top, parskip=0pt, aboveskip=0pt, justification=raggedright, labelseparator=none, farskip=0pt, nearskip=2pt, margin=0pt, captionskip=-15pt, font=normalsize, singlelinecheck=false}
	\includegraphics[width= 8.6cm]{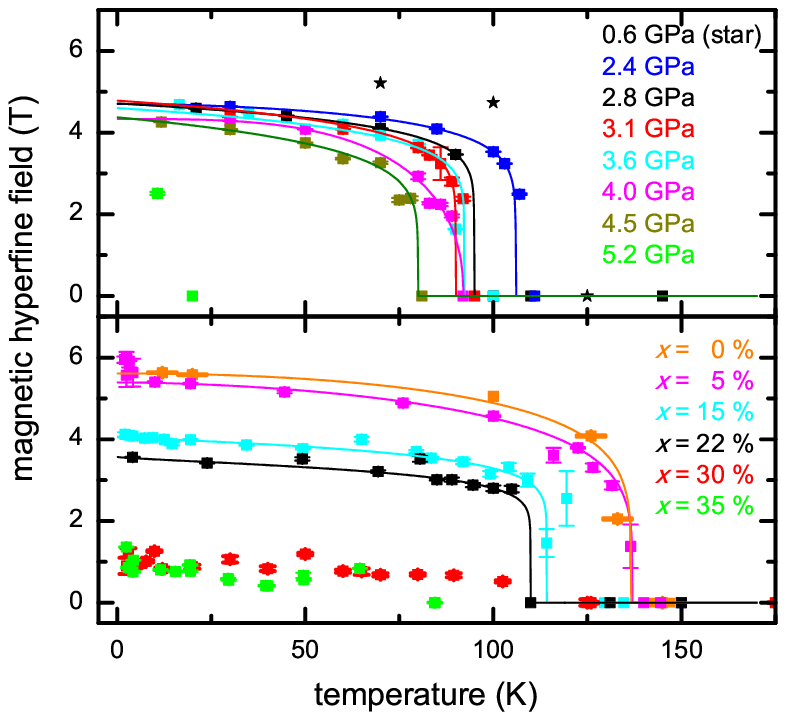} \label{fig:B-p-x}
	\caption{Temperature dependence of the magnetic hyperfine field as a function of \textit{p} (top) and \textit{x} (bottom) in $\mathrm{CeFeAsO}$. Lines are calculated with Eq.\;{}\ref{eq:OP}.}
	\label{fig:Bhf}
\end{figure}

The temperature dependence of the magnetic hyperfine field, \bhf(\textit{T}), as a function of \textit{x} and \textit{p} is shown in Fig.~\ref{fig:Bhf}.
\bhf(\textit{T}) was analyzed using an order parameter fit of the form 
\begin{equation}
	B_{\mathrm{hf}}(T) = B_{\mathrm{hf}}(T=0) \left [ 1-\left( \frac{T}{T_{\mathrm{N}}} \right )^{\alpha} \right ]^{\beta}
	\label{eq:OP}
\end{equation}
at temperatures above the magnetic Ce ordering.
The results are shown in Tab.\;{}\ref{tab:OP}.
\begin{table}[htbp]
\caption{Exponents $\alpha$ and $\beta$ obtained by analyzing the temperature dependence of the magnetic hyperfine field applying Eq.\;{}\ref{eq:OP} to temperatures above the Ce magnetic order. To determine the critical exponent $\beta_{\mathrm{c}}=\;{}\beta(\alpha\;{}=\;{}1)$, Eq.\;{}\ref{eq:OP} was applied in the vicinity of the phase transition and with $\alpha$\;{}=\;{}1.}
	\centering
		\begin{tabular}{ccccc}
		\hline\hline
		\textit{p} / GPa 	&	\textit{x} / \% 	&	$\alpha$	&	$\beta$	&	$\beta_{\mathrm{c}}$ \\
		0.6	&			&					&					&	0.09(4)	\\
		2.4	&			&	3.0(2)	&	0.18(1)	&	0.17(1)	\\
		2.8	&			&	2.3(1)	&	0.18(1)	&	0.14(1)	\\
		3.1	&			&	2(1)		&	0.17(6)	&	0.12(1)	\\	
		3.6	&			&	1.8(2)	&	0.15(1)	&	0.12(1)	\\
		4.0	&			&	3.3(4)	&	0.42(6)	&	0.13(3)	\\
		4.5	&			&	1.2(6)	&	0.18(8)	&	0.13(4)	\\
				&	0		&	2.6(2)	&	0.25(1)	&	0.17(1)	\\
				&	5		&	1.7(2)	&	0.02(1)	&	0.14(1)	\\
				&	15	&	0.6(4)	&	0.09(3)	&	0.10(1)	\\
				&	22	&	0.7(6)	&	0.09(5)	&	0.10(3)	\\
\hline\hline
\end{tabular}
	\label{tab:OP}
\end{table}

Both the onset of the magnetic order as well as the saturated magnetic hyperfine field at lowest temperatures are continuously suppressed as a function of \textit{x}.
For \textit{x}\;{}=\;{}40\;{}\% no magnetic order was observed which is consistent with results from other methods.\cite{PhysRevB.86.020501, PhysRevLett.104.017204, PhysRevB.81.134422}
For \textit{x}\;{}=\;{}5\;{}\% an increase of the magnetic hyperfine field from 5.40(2)\;{} above 4\;{}K to 5.95(8)\;{}T below 4\;{}K is observed due to the antiferromagnetic ordering of the Ce 4\textit{f} electrons.\cite{PhysRevB.86.020501}
This transferred magnetic hyperfine field was also observed in the unsubstituted compound $\mathrm{CeFeAsO}$ where an increase by 0.9\;{}T was measured.\cite{1367-2630-11-2-025011, PhysRevB.80.094524, 0953-2048-25-8-084009}
Increasing \textit{x} to 15\;{}\% or above leads to a full suppression of this transfer. 

The saturated magnetic hyperfine field is suppressed with increasing applied pressure (Fig.~\ref{fig:PD-B-x-p}).
Between 0 and 4.5\;{}GPa the saturated magnetic hyperfine field is reduced by $\sim$\;{}24\;{}\% followed by an abrupt suppression to zero.
Between 5.2 and 14\;{}GPa no magnetic order was found down to 16\;{}K.

\begin{figure}[htbp]
	\centering
	\includegraphics[width= \columnwidth]{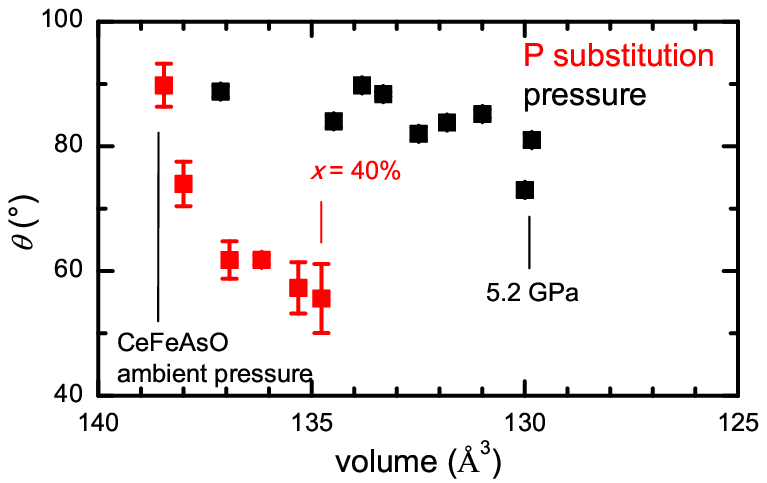} 	
	\caption{Azimuth angle $\theta$ between the principal axis of the EFG and the Fe magnetic hyperfine field as a function of the P-substitution level \textit{x} (square) and hydrostatic pressure (star).}
	\label{fig:theta}
\end{figure}

The azimuth angle $\theta$ between the principal axis of the EFG, which is parallel to the crystallographic \textit{c}-axis, and the magnetic hyperfine field at lowest observed temperatures is shown in Fig.~\ref{fig:theta} as a function of \textit{x} and \textit{p}.
For $\mathrm{CeFeAsO}$ at ambient conditions an angle of $\theta$\;{}=\;{}90$^{\circ}$ was obtained.
Therefore, the Fe magnetic moments are located in the crystallographic \textit{ab}-plane consistent with neutron scattering experiments.\cite{nmat2315}
Upon the application of pressure a small tilting of 10$^{\circ}$ out of the crystallographic \textit{ab}-plane at 4.5\;{}GPa is observed.
In contrast, $\theta$ increases up to 56(6)$^{\circ}$ as a result of P substitution.

\subsection{Isomer and chemical shift in CeFeAs$_{1-x}$P$_x$O}

\begin{figure}[htbp]
	\centering
	\includegraphics[width= \columnwidth]{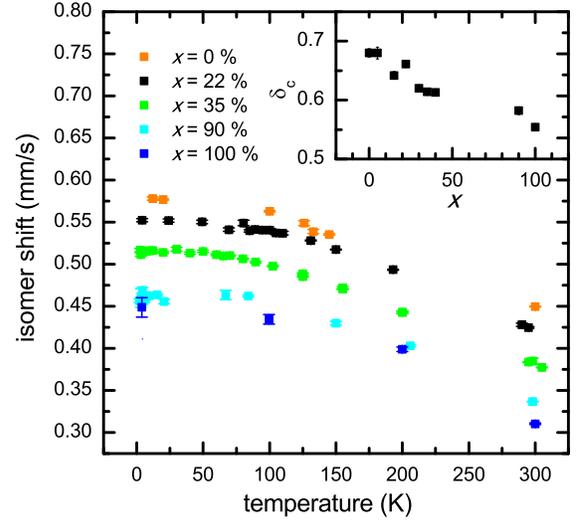} 	
	\caption{Temperature dependence of the isomer shift of \textit{x}\;{}=\;{}0, 22, 35, 90, and 100\;{}\% (\textit{x}\;{}=\;{}5, 15, and 40 \;{}\% are omitted for the sake of clarity). In the inset the chemical shift $\delta_{\mathrm{c}}$ in mm/s as a function of \textit{x} in \% is shown.}
	\label{fig:IS}
\end{figure}

The temperature dependence of the isomer shift, $\delta(T)$, for selected P-substitution levels is shown in Fig.~\ref{fig:IS}.
$\delta(T)$ is given by
\begin{equation}
	\delta (T) = \delta_{\mathrm{c}} + \delta_{\mathrm{R}} (T),	\label{eq:cs} 
\end{equation}
where $\delta_{\mathrm{c}}$ denotes the temperature-independent chemical shift.
$\delta_{\mathrm{R}} (T)$ is the temperature-dependent contribution due to the second-order Doppler shift and was analyzed in the Debye approximation:
\begin{align}
	\delta_{\mathrm{R}} (T) = &-\frac{9}{16} \frac{k_\mathrm{B}}{M_{\mathrm{Fe}}~\mathrm{c}} \\\nonumber
	 &\times \left [ \theta_{\mathrm{M}} + 8T \left ( \frac{T}{\theta_{\mathrm{M}}} \right )^3 \int_0^{\theta_{\mathrm{M}} / T} \frac{x^3}{\mathrm{e}^x -1}\mathrm{d}x \right ]
\end{align}
with \textit{M}$_{\mathrm{Fe}}$ being the mass of the resonant $^{57}$Fe nucleus and \td~ denotes the M\"ossbauer temperature.
\td~ can be interpreted as the Debye temperature of the Fe nucleus.
By fixing \textit{M}$_{\mathrm{Fe}}$ to its nuclear value of 56.93\;{}a.u., \td~ and $\delta_{\mathrm{c}}$ were calculated.
The obtained results are shown in Tab. \ref{tab:td}.
\begin{table}[htbp]
\caption{M\"ossbauer temperature \td\,{} and chemical shift $\delta_{\mathrm{c}}$ obtained by applying Eq. \ref{eq:cs} to the temperature dependence of the isomer shift in CeFeAs$_{1-x}$P$_x$O.}
\centering
		\begin{tabular}{ccc}
		\hline\hline
		\textit{x} / \% 	&	\td / K 	&	$\delta_{\mathrm{c}}$ / mm/s \\
		0		&	381(32)	&	0.680(7)	\\
				&	377(5)\cite{PhysRevB.87.064302}	& \\
		5		&	445(53)	&	0.68(1)	\\
		15	& 342(26)	&	0.642(6)\\
		22	&	401(13)	&	0.661(3) \\
		30	&	385(5)	&	0.620(1)	\\
		35	& 360(5)	&	0.614(1) \\\
		40	&	401(12)	&	0.613(3)	\\
		90	&	438(23)	&	0.582(6)	\\
		100	&	448(31)\cite{JPSJ.81.064714}	&	0.55\cite{JPSJ.81.064714}\\		
\hline\hline
\end{tabular}
	\label{tab:td}
\end{table}

We found an increase of \td\,{} from $\mathrm{CeFeAsO}$ to CeFePO upon As$\rightarrow$P substitution while $\delta_{\mathrm{c}}$ decreases.

\section{discussion}
\label{sec:discussion}
To reveal the differences in the electronic structure between hydrostatic pressure and P substitution, following Ref.\cite{PhysRevB.88.054428}, we compared the obtained electronic hyperfine parameter as a function of the unit cell volume.
The relation between the unit cell volume and \textit{x} and \textit{p} is shown in Fig. \ref{fig:xrd}.

In the paramagnetic phase $\mathrm{CeFeAsO}$ has a \vzz \,{} of close to zero indicating a nearly spherical charge distribution around the Fe nucleus while CeFePO has a \vzz \,{} of 9.3(2)\;{}V/\AA$^2$ indicating a deviation from a spherical charge distribution.
This is consistent with reported results from neutron scattering experiments.\cite{PhysRevLett.104.017204}
They found a continuous reduction of the size of the \textit{Pn}-Fe-\textit{Pn} block, with \textit{Pn}\;{}=\;{}As/P, as well as a continuous reduction of the Fe-\textit{Pn} distance.
As a consequence the Fe-\textit{Pn}-Fe tetrahedra angle increases from $\sim$ 112.2$^{\circ}$ to $\sim$ 114.6$^{\circ}$ for \textit{x} = 0 and 43 \%, respectively.
Thus the angle continuously deviates from the ideal value of 109.47$^{\circ}$ with increasing \textit{x}.
This continuous change in the Fe\textit{Pn} block properties explains the increase in \vzz \,{} from $\mathrm{CeFeAsO}$ to CeFePO but cannot explain the maximum at intermediate \textit{x}.
We attribute this maximum to the disorder induced by the substitution which we expect to be strongest at \textit{x}\;{}$\sim$\;{}50\;{}\%.

In contrast, the value of \vzz \,{} in the paramagnetic phase increases only slightly and monotonically as a function of hydrostatic pressure.
This indicates a slight deviation from the spherical charge distribution around the Fe nucleus with increasing pressure.
As it was shown in Fig. \ref{fig:lattice} the As-Fe-As layer remains robust against the application of hydrostatic pressure.
Reported high temperature Fe-As-Fe angles for CeFeAsO are 112.6(1)$^{\circ}$.\cite{nmat2315, PhysRevB.82.134514}
We observed a minor reduction of the Fe-As-Fe angle to $\sim$ 112.2$^{\circ}$ at 5.3 GPa.
Due to the tetragonal symmetry a reduction of the Fe-As distance will not increase \vzz.
The crystallographic parameters which are significantly changing in the investigated pressure region are the \textit{c}/\textit{a} ratio and the Ce-As distance and thus the As-Fe-As and Ce-O-Ce block distance.
However, both crystallographic parameters are expected to have only a minor influence on \vzz, in contrast to the As-Fe-As block size in the P-substituted compound.\cite{PhysRevLett.104.017204}
Eventually the only small increase of \vzz \,{} reflects the robustness of the As-Fe-As layer against hydrostatic pressure and the only minor changes in the Fe-As-Fe angle. 

At the magnetic phase transition temperature \vzz \,{} abruptly increases and remains constant within error bars down to lowest temperatures.
We obtained that \vzz \,{} remains constant within error bars in the magnetic phase at all investigated pressures.
We found that the abrupt increase in \vzz \,{} at the magnetic phase transition is suppressed with increasing external pressure similar to the magnetic hyperfine field.
For $\mathrm{CeFeAsO}$ a splitting in temperature between the structural and magnetic phase transition was reported.\cite{PhysRevB.81.134525, PhysRevLett.104.017204}
We observed no change in \vzz \,{} at the structural phase transition.
For comparison: in FeSe, where a tetragonal-to-orthorhombic phase transition without a coinciding magnetic order occurs, similarly no change in \vzz \,{} was observed.\cite{Blachowski20101}
This indicates that the magnetic phase transition causes a redistribution of the electronic charge and hence changes \vzz \,{} while the changes due to the structural phase transition are negligible.
This result also explains why the abrupt increase of \vzz \,{} at the magnetic phase transition and the magnetic hyperfine field are equally suppressed.

\begin{figure}[htbp]
	\centering
	\includegraphics[width= \columnwidth]{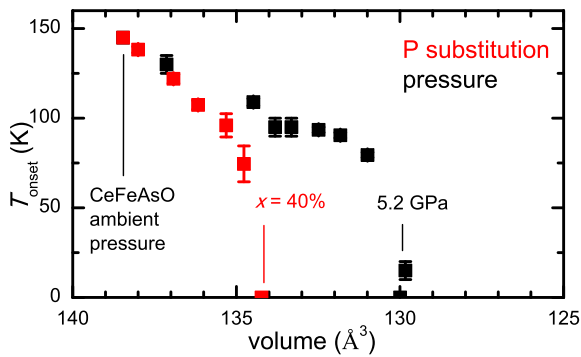} 	
	\caption{Onset temperature of the Fe magnetic order as a function of hydrostatic pressure (black) and P substitution (red). \ton \,{} is suppressed in a qualitatively similar way as a function of hydrostatic pressure and P substitution but more effective by the latter.}
	\label{fig:PD-T-x-p} 
\end{figure}

\ton~ shows qualitatively similar behavior for increasing \textit{x} and \textit{p} and is shown in Fig. \ref{fig:PD-T-x-p}.
\ton~ is continuously reduced until \textit{x}\;{}$\sim$\;{}30\;{}\% and \textit{p}\;{}$\sim$\;{}4.5\;{}GPa followed by a sharp suppression to zero at \textit{x}\;{}$\sim$\;{}40\;{}\% and \textit{p}\;{}$\sim$\;{}5.2\;{}GPa.
No magnetic order was observed at higher values.
For the P substitution series it is consistent with neutron scattering experiments where no magnetic order was found at \textit{x}\;{}$\sim$\;{}37\;{}\%.\cite{PhysRevLett.104.017204}
The phase transition region $\Delta$\textit{T}\;{}=\;{}\ton$-$\;{}\tfull~increases with increased \textit{x}.
We attribute this to the fact that the P substitution results in local P distributions and hence a distribution of magnetic ordering temperatures.
In contrast $\Delta$\textit{T} remains constant within error bars for all applied pressures.
This supports that the increase in $\Delta$\textit{T} is caused by the disorder due to the P substitution.

The temperature dependence of the magnetic hyperfine field was analyzed using Eq.\;{}\ref{eq:OP}.
We found a second-order phase transition in the magnetic hyperfine field for all investigated pressures and P-substitution levels consistent with published results for $\mathrm{CeFeAsO}$.\cite{PhysRevB.81.134525}
A critical exponent \bc\,{} of 0.17(1) was obtained in $\mathrm{CeFeAsO}$.
Both the application of pressure and P substitution result in a reduction of \bc\,{} in direction of the two-dimensional Ising universality class (\bc\;{}=\;{}0.125).
This behavior indicates an increase of the two-dimensionality of the magnetic order.
Published M\"ossbauer data suggest that the critical exponent of the magnetic hyperfine field is of similar value in LaFeAsO (0.2(1)) and PrFeAsO (0.19(2)).\cite{1367-2630-11-2-025011}
A 2D Ising critical exponent was also found in BaFe$_2$As$_2$,\cite{PhysRevB.79.184519, PhysRevB.81.014501} SrFe$_2$As$_2$,\cite{PhysRevB.78.140504} Ba$_{0.75}$Na$_{0.25}$Fe$_2$As$_2$,\cite{2012arXiv1210.6881M} and Ca$_{0.65}$Na$_{0.35}$Fe$_2$As$_2$.\cite{PhysRevB.92.134511}
The reduction in \bc\,{} and therefore in the dimensionality due to chemical pressure was also observed in Ca$_{1-x}$Na$_{x}$Fe$_2$As$_2$ where a crossover from three- to two-dimensional Ising behavior from 50\;{}\% to 30\;{}\% Na-substitution level was found.\cite{PhysRevB.92.134511}

\begin{figure}[htbp]
	\centering
	\includegraphics[width= \columnwidth]{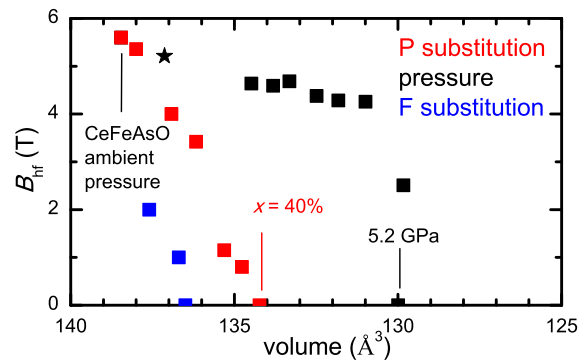} 	
	\caption{Low-temperature saturated magnetic hyperfine field (above the Ce ordering temperature) as a function of hydrostatic pressure (black) and P substitution (red). The black star data point was taken at 70\;{}K and 0.6\;{}GPa. P substitution results in a continuously suppression of the magnetic hyperfine field to zero at \textit{x} = 40 \%. In contrast, the application of hydrostatic pressure results in a reduction of the magnetic hyperfine field by 24 \% at 4.5 GPa followed by an abrupt reduction to zero at 5.2 GPa. CeFeAsO$_{1-y}$F${_y}$ data (blue) taken from Ref.\cite{2012arXiv1210.6959M}}
	\label{fig:PD-B-x-p} 
\end{figure}
The low-temperature saturated magnetic hyperfine field (above the Ce ordering temperature) as a function of \textit{x} and \textit{p} is shown in Fig.~\ref{fig:PD-B-x-p}.
It is continuously reduced to zero with increasing \textit{x}.
This behavior as a function of \textit{x} is similar to that of the Fe magnetic moment and orthorhombicity.\cite{PhysRevLett.104.017204}
In contrast, the saturated magnetic hyperfine field is reduced by 24\;{}\% between 0 and 4.5\;{}GPa followed by an abrupt suppression to zero above 5.2\;{}GPa showing a behavior similar to \ton.

\begin{figure}[htbp]
	\centering
	\includegraphics[width= \columnwidth]{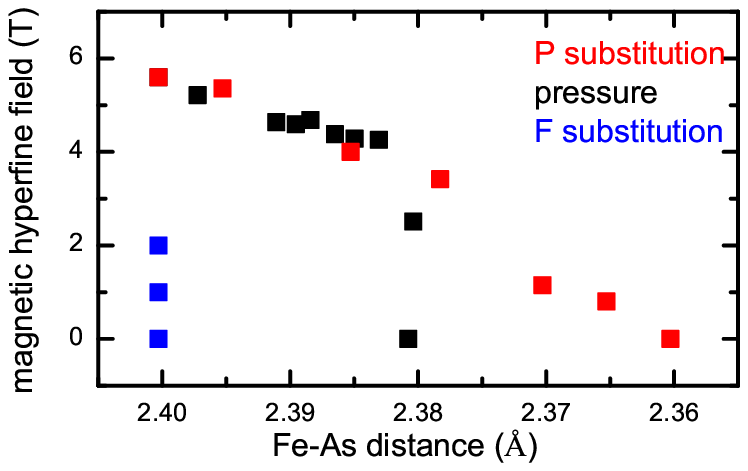} 	
	\caption{Magnetic hyperfine field as a function of the Fe-As distance. Fe-As distance as a function of the P-substitution level is taken from Ref.\cite{PhysRevLett.104.017204}. CeFeAsO$_{1-y}$F${_y}$ data (blue) taken from Ref.\cite{2012arXiv1210.6959M}}
	\label{fig:Bhf-dFeAs} 
\end{figure}

It was shown that the Fe magnetic moment is proportional to the Fe-As distance and vanishes for distances smaller than 2.36 \AA.\cite{PhysRevLett.104.017204, PhysRevLett.101.126401, PhysRevB.67.155421}
The Fe magnetic moment  is not directly accessible by M\"ossbauer spectroscopy which measures the magnetic hyperfine field.
Theoretical calculations on BaFe$_2$As$_2$ have shown that the conversion factor between the Fe magnetic moment and the magnetic hyperfine field changes with chemical substitution.\cite{PhysRevB.94.214508}
The changes in the conversion factor are severe for electron and hole doping but are below 3 \% for P substitution.\cite{PhysRevB.94.214508}
Unfortunately no calculations for hydrostatic pressure were performed but it is more likely that the conversion factor exhibits only minor changes in the case of hydrostatic pressure.\cite{PhysRevB.94.214508}
Therefore we treat the conversion factor between Fe magnetic moment and magnetic hyperfine field as constant in our work.
The saturated low-temperature magnetic hyperfine field as a function of the Fe-As distance is shown in Fig. \ref{fig:Bhf-dFeAs}.

At this point we want to emphasize that our XRD measurements were conducted at room temperature while the reported neutron scattering data was obtained at 1.8 to 8 K.\cite{PhysRevLett.104.017204}
However, the Fe-As distance is nearly temperature independent with changes below 0.002 \AA \,{} between room temperature\cite{PhysRevB.82.134514} and 1.8 K\cite{nmat2315} in CeFeAsO and therefore we assume that this also the case in the P-substituted compounds and under hydrostatic pressure.\cite{PhysRevLett.104.017204}
The Fe-As distance continuously decreases below the threshold value 2.36 \AA \,{}for \textit{x} $\sim$ 37 \%.\cite{PhysRevLett.104.017204}
The reduction of the magnetic hyperfine field as a function of applied hydrostatic pressure shows the qualitatively same behavior above 2.38 \AA.
This result supports that the Fe magnetic moment is somewhat  proportional to the Fe-$Pn$ distance which determines the hybridization strength of the Fe 3\textit{d} and $Pn$ \textit{p} electrons.
The observation of a purely paramagnetic phase at 5.3 GPa with a Fe-As distance of $\sim$ 2.38 \AA \,{} implies that the \textit{dp} hybridization is not the only mechanism controlling the Fe magnetic moment.
This is supported by measurements in CeFeAsO$_{1-y}$F$_y$ where a reduction of the magnetic moment to zero with increasing F-substitution level while having a nearly constant Fe-As distance of $\sim$ 2.405 \AA \,{} was observed.\cite{2012arXiv1210.6959M, nmat2315}
That the Fe-As-Fe angle increases with increasing \textit{x} but decreases with increasing \textit{p} may also play a role.
Additionally, the strong suppression to zero occurs between 4.5 and 5.2 GPa.
In this pressure region a maximum in the magnetic phase transition temperature of the Ce 4\textit{f} electrons was reported.\cite{PhysRevB.83.094528}
In CeFeAs$_{0.78}$P$_{0.22}$ it was observed that the Ce 4\textit{f} magnetic ordering temperature has a maximum at 1.95 GPa where the magnetic order changes from anti- to ferromagnetic.\cite{PhysRevB.86.134523}
The resulting question is if the anti- to ferromagnetic transition also occurs in CeFeAsO between 4.5 and 5.2 GP and if the Ce ferromagnetic order strongly suppresses the Fe magnetic order.

To derive an explanation of our obtained results we want to compare them with results from density-functional theory (DFT) in the La-1111 compounds and add the additional Ce 4\textit{f} component later.
Comparing LaFeAsO and LaFePO at ambient pressure shows that the Fermi surfaces (FS) are comparable with three hole pockets at $\Gamma$ and two electron pockets at \textit{M}.\cite{PhysRevB.78.064518,PhysRevLett.100.237003}
The states at the Fermi level are mostly of Fe 3\textit{d} character.\cite{1742-6596-150-5-052090}
The difference is that one of the hole pockets is of 2D character in LaFeAsO and of 3D character LaFePO.
This indicates better nesting in LaFeAsO and hence magnetic ordering than in LaFePO.
It was shown that the FS topology is sensitive to the local Fe-\textit{Pn} arrangement, namely the Fe-\textit{Pn} distance and the corresponding tetrahedra angle.\cite{PhysRevB.78.024521, PhysRevB.78.064518}
This sensitivity is due to the hybridization of the Fe 3\textit{d} and \textit{Pn} \textit{p} states.\cite{PhysRevB.78.064518}
Our experiments as well as published results show that the Fe\textit{Pn} distance decreases with increasing pressure\cite{1.4904954} and As$\rightarrow$P substitution\cite{PhysRevLett.104.017204, nmat3120} in 1111 compounds.
An decreased Fe\textit{Pn} distance results in an enhanced Fe\textit{Pn} hybridization.\cite{PhysRevB.78.064518}

DFT calculations in the paramagnetic state found that the density of states (DOS) at the Fermi level is smaller for LaFePO than for LaFeAsO.\cite{1742-6596-150-5-052090}
DFT calculations on LaFeAsO under pressure in the magnetically order phase show that the FS topology is rather robust which implies that the nesting condition remains intact.\cite{PhysRevB.79.024509}
In this study it was also found that the DOS at the Fermi level decreases with increasing pressure in LaFeAsO.\cite{PhysRevB.79.024509}
This is consistent with the observation that the DOS at Fermi level is somewhat proportional to magnitude of the magnetic order parameter.\cite{PhysRevB.78.094511, PhysRevB.78.085104, PhysRevLett.101.047001}
In summary DFT calculations in LaFeAsO indicate that As$\rightarrow$P substitution changes the dimensionality of one hole pocket from 2D to 3D and thus weakens the nesting properties while the FS remains robust under the application of pressure.
This is consistent with observations in BaFe$_2$(As$_{1-x}$P$_x$)$_2$ that the isovalent substitution changes the FS similar to charge doping.\cite{PhysRevLett.104.057008}
This explains the differences in the magnetic properties on a qualitative level if we take into account that the application of $\approx$\;{}8\;{}GPa in LaFeAsO has the same effect on the unit cell volume as the transition from LaFeAsO to LaFePO with the former showing magnetic order and the latter being paramagnetic.

If we replace La by Ce and thus adding one 4\textit{f} electron the discussion follows the same arguments.
Published calculations in the paramagnetic state used DFT + dynamical mean-field theory (DMFT) to account for the additional 4\textit{f} correlations.\cite{0295-5075-84-3-37006}
Similar to the La-1111 compounds the DOS at the Fermi level is mostly of Fe 3\textit{d} character.\cite{0295-5075-84-3-37006}
It was found that CeFePO has a smaller DOS at the Fermi level than $\mathrm{CeFeAsO}$.\cite{0295-5075-84-3-37006}
An applied pressure of $\sim$\;{}5\;{}GPa on $\mathrm{CeFeAsO}$ yields the same DOS at the Fermi level as CeFePO.\cite{0295-5075-84-3-37006}
In addition the hybridization of the Fe 3\textit{d} and Ce 4\textit{f} states in $\mathrm{CeFeAsO}$ is much smaller than in CeFePO.\cite{0295-5075-84-3-37006}
This is consistent with angle-resolved photo emission spectroscopy measurements in the P-substitution series where a change in the FS and an increase in the 3\textit{d}-4\textit{f} hybridization from \textit{x} = 30 to 100 \% was observed.\cite{PhysRevLett.104.096402, PhysRevB.86.020506}
The application of $\sim$\;{}10\;{}GPa on $\mathrm{CeFeAsO}$ results in a hybridization similar to CeFePO.

The transferred magnetic hyperfine field due to the magnetic order of the Ce 4\textit{f} electrons is reduced from \textit{x}\;{}=\;{}0 to 5\;{}\% and was not observed for 15\;{}\% and higher \textit{x}.
This implies that the ordered moment of the Ce 4\textit{f} electrons is strongly reduced as a function of \textit{x}.
This is consistent with reported results that the Ce 4\textit{f} ordering temperature is independent from \textit{x} but the ordered moment is rapidly suppressed with increased \textit{x}.\cite{PhysRevLett.104.017204}

The \textit{x} dependence of the chemical shift is shown in the inset of Fig.~\ref{fig:IS} and in Tab. \ref{tab:td}.
We found a reduction of the chemical shift $\delta_{\mathrm{c}}$ with increasing \textit{x}.
A reduction in $\delta_{\mathrm{c}}$ corresponds to an increase in the electron density at the Fe nucleus.
It was reported that the Fe-\textit{Pn} distance decreases with increasing \textit{x}.\cite{PhysRevLett.104.017204}
This may increase the hybridization of the Fe 3\textit{d} and \textit{Pn} \textit{p} valence electrons.\cite{PhysRevB.78.024521, PhysRevLett.101.126401}
The hybridization of the Fe 3\textit{d} with the Ce 4\textit{f} electrons as it was observed in CeFePO may also play a role.\cite{PhysRevLett.104.096402, PhysRevB.86.020506}
As a consequence the shielding of the Fe 4\textit{s} electrons by the Fe 3\textit{d} is reduced resulting in an increased electron density at the nucleus.

\section{summary and conclusion}
\label{sec:summary}
In summary, we performed in-house and synchrotron M\"ossbauer spectroscopy experiments on CeFeAs$_{1-x}$P$_x$O powder and on $\mathrm{CeFeAsO}$ single crystals, the latter under hydrostatic pressure and provide an updated microscopical phase diagrams in combination with XRD measurements.
We found a qualitatively similar suppression of the onset temperature of the Fe magnetic order as a function of \textit{x} and \textit{p}.
In contrast, the low-temperature saturated magnetic hyperfine field is continuously suppressed to zero at \textit{x} = 40 \% while it is reduced by 24 \% between 0 and 4.5 GPa followed by an abrupt suppression to  zero.
Above \textit{x} = 40 \% and \textit{p} = 5.2 GPa we observed no Fe magnetic order.
We found that the magnetic hyperfine field is proportional to the Fe-As distance above 2.38 \AA \,{} for both hydrostatic pressure and P substitution.
The observation of a paramagnetic phase for a Fe-As distance of 2.38 \AA \,{} which is above the threshold value of 2.36 \AA \,{} implies that the magnetic moment is not only controlled by the \textit{dp} hybridization.
Our study suggests that the size of the Fe magnetic moment is the result of a delicate interplay of the Fe 3\textit{d}, $Pn$ \textit{p}, and Ce 4\textit{f} electrons and goes beyond the Fe 3\textit{d} $Pn$ \textit{p} hybridization.
We conclude that hydrostatic pressure change both the crystallographic and electronic properties of the system differently than P substitution.

\acknowledgments
Part of this work was funded by the Deutsche Forschungsgemeinschaft (DFG, German Research Foundation) -- MA 7362/1-1, JE 748/1, WU595/3-3, BU887/15-1, and the research training group GRK-1621.
This research used resources of the Advanced Photon Source, a U.S. Department of Energy (DOE) Office of Science User Facility operated for the DOE Office of Science by Argonne National Laboratory under Contract No. DE-AC02-06CH11357.
Parts of this work were performed at GeoSoilEnviroCARS (Sector 13), Partnership for Extreme Crystallography program (PX$^2$), Advanced Photon Source (APS), and Argonne National Laboratory. GeoSoilEnviroCARS is supported by the National Science Foundation-Earth Sciences (EAR-1634415) and Department of Energy-Geosciences (DE-FG02-94ER14466). The COMPRES-GSECARS gas loading system and the PX$^2$ program are supported by COMPRES under NSF Cooperative Agreement EAR-1661511.
We thank S. Tkachev for help with the Ne loading of the DAC.

\bibliography{Ce1111}

\newpage
\section{APPENDIX}
Supplement: Numeric values of the presented figures as well as the crystallographic data of CeFeAsO under pressure:

\begin{table*}[htbp]
\caption{Crystallographic parameters as a function of pressure at room temperature}
\begin{center}
\begin{tabular}{ccccccc}
\hline\hline
\textit{p} / GPa & \textit{a} / \AA & \textit{c} / \AA & \textit{z}(Ce)	&	\textit{z}(As)	&	\textit{R} / \%	\textit{wR} / \%\\
5.3&	3.9452(2) &	8.3388(6)	&	0.1481(5)	&	0.6590(9)	&	5.77	&	7.99 \\
4.55&	3.9535(2)	& 8.3771(6)	&	0.1476(6)	&	0.6583(9)	&	6.19	&	8.88	\\
3.92&	3.9614(2)	& 8.4100(6)	&	0.1463(5)	&	0.6576(8)	&	6.32	&	9.03 \\
3.53&	3.9665(2)	& 8.4341(5) &	0.1457(5)	&	0.6582(9)	&	5.55	&	7.7 \\
3.2	& 3.9711(2)	& 8.4563(6)	&	0.1450(5)	&	0.6558(8)	&	5.95	&	8.12 \\
2.9	& 3.9743(2)	& 8.4720(5)	&	0.1447(5)	&	0.6569(8)	&	5.48	&	7.53 \\
2.38&	3.9771(2)	& 8.4868(5)	&	0.1444(5)	&	0.6564(8)	&	5.5	&	7.65 \\
1.85&	3.9836(2)	& 8.5221(6)	&	0.1434(5)	&	0.6558(8)	&	5.68	&	8.03 \\
1.57&	3.9886(2) & 8.5469(6)	&	0.1426(5)	&	0.6540(8)	&	5.01	&	8.19 \\
\hline\hline
\end{tabular}
\end{center}
\label{lattice}
\end{table*}

\begin{table*}[htbp]
\caption{Fig. 5: \vzz~ as a function of \textit{x} in the paramagnetic and magnetically ordered temperature regime}
\begin{center}
\begin{tabular}{ccccc}
\hline\hline
\textit{x} / \%	&	\textit{T} / K	&	\vzz~/ V/\AA$^2$	&	\textit{T} / K	&	\vzz~/ V/\AA$^2$ \\\hline
0	&	300	&	1.3(1.1)	&	12	&	7.8(4)	\\
5	&	298	&	0(0.5)	&	10	&	6.7(8) \\
15&	305	& 9.7(2)	&	4.2		&	14.2(1.1)\\
22&	295	& 10.2(2)	& 4.1	& 14.2(5) \\
30&	305	& 12.78(2)	&3	&14.5(1.9)\\
35&	305	& 12.95(2)	& 4.2	& 13(1.5) \\
40&	305	& 12.95(6)	& 2	& 14.8(2) \\
90&	298	& 11.00(5)  & 2.2	& 	13.2(4) \\
100&	300 & 	9.3(2) & 	4.2	& 12.41(8) \\
\hline\hline
\end{tabular}
\end{center}
\label{Vzz}
\end{table*}

\begin{table*}[htbp]
\caption{Fig. 8: low-temperature values of the azimuth angle $\theta$ between the principal axis of the electric field gradient and the magnetic hyperfine field}
\begin{center}
\begin{tabular}{cccc}
\hline\hline
\textit{p} / GPa	&	$\theta$ / $^{\circ}$	&	\textit{x} / \%		&	$\theta$ / $^{\circ}$ \\\hline
0.5 & 88.8(9) & 0 & 90.2(3.5) \\
2.65 & 84(1)	& 5 & 106.0(3.6) \\ 
2.36 & 89.8(4) & 15 & 118(3) \\
3.05 & 88.4(4) & 22 & 61.8(7) \\ 
2.29 & 82(1) & 30 & 123(4) \\
3.54 & 83.8(9) & 35 & 55.6(5.5) \\
3.87 & 85.2(7) & &  \\ 
4.4 & 81(1) &  &  \\ 
5.16 & 73(1)&  &  \\ 
\hline\hline
\end{tabular}
\end{center}
\label{theta}
\end{table*}

\begin{table*}[htbp]
\caption{Fig. 10}
	\centering
		\begin{tabular}{ccc}
		\hline\hline
		\textit{p} / GPa 	&	\textit{x} / \% 	&	\textit{T}$_{\mathrm{N}}^{\mathrm{onset}}$ / K \\
		
0.8(1)&&	130(5) \\
2.4(1)&&	109(2) \\
2.8(1)&&	95(5) \\
3.1(1)&&	95(5)	\\
3.6(1)&&	93.5(1.5)	\\
4.0(1)&&	90.5(1.5)	\\
4.5(1)&&	79.5(1.5)	\\
5.2(1)&&	15(5)	\\
5.1(1)&&		0	\\
&5&		138.25(1.75)	\\
&15&	122(2.5)	\\
&22&	107.4(2.5)\\
&30&	96(6.5)\\
&35&	74.5(10)\\
&40	&0\\
\hline\hline
\end{tabular}
	\label{tab:Fig1a}
\end{table*}

\begin{table*}[htbp]
\caption{Fig. 11, errors in \textit{B}$_{\mathrm{hf}}$ are below 1\;{}\%}
	\centering
		\begin{tabular}{ccc}
		\hline\hline
		\textit{p} / GPa 	&	\textit{x} / \% 	&	\textit{B}$_{\mathrm{hf}}$ / T\\
		0	&&	5.6 \\
0.5(1)	&&		5.2 \\
2.3(1)	&&		4.6 \\
2.7(1)	&&		4.6 \\
3.1(1)	&&		4.7 \\
3.5(1)	&&		4.4 \\
3.9(1)	&&		4.3 \\
4.5(1)	&&		4.3 \\
5.2(1)	&&		2.5 \\
5.1(1)	&&		0 \\
&0&	5.6 \\
&5&	5.4 \\
&15&	4 \\
&22&	3.4 \\
&30&	1.2 \\
&35&	0.8 \\
&40&	0 \\
\hline\hline
\end{tabular}
	\label{tab:Fig1b}
\end{table*}
\end{document}